\documentclass{jpsj-suppl}
\usepackage{txfonts}
\usepackage[english]{babel}
\usepackage[T1]{fontenc}
\usepackage{bm}

\title{The EOS of neutron matter and the effect of $\Lambda$ hyperons to neutron star structure}

\author{Stefano \textsc{Gandolfi}$^{1}$ and Diego \textsc{Lonardoni}$^{1,2}$}

\inst{$^{1}$Theoretical Division, Los Alamos National Laboratory, Los Alamos, New Mexico 87545, USA \\
$^{2}$National Superconducting Cyclotron Laboratory, Michigan State University, 640 South Shaw Lane, East Lansing, Michigan 48824, USA}

\recdate{\today}

\abst{The structure of neutron stars is determined by
the equation of state of the matter inside the star, which relies on the 
knowledge of nuclear interactions. 
While radii of neutron stars mostly depend on the equation of state of neutron
matter at nuclear densities, their maximum mass can be drastically affected
by the appearance of hyperons at higher densities in the inner core of the star. 
We summarize recent quantum Monte Carlo results on the calculation of the 
equation of state of neutron matter at nuclear and higher densities. We report
about the development of realistic hyperon-nucleon interactions based on the available 
experimental data for light- and medium-heavy hypernuclei and on the effect of 
$\Lambda$ hyperons to the neutron star structure.}

\kword{neutron stars, neutron matter, hyperons, hypernuclei, hyperon puzzle}

\begin{document}
\maketitle

\section{Introduction}
Neutron stars are the most compact and dense objects in the universe,
with typical masses $M\sim1.4~M_\odot$ and radii $R\sim10$~km.  Their
central densities can be several times larger than nuclear saturation
density, $\rho_0=0.16~\text{fm}^{-3}$, corresponding to the central
density of heavy atomic nuclei. Because at such high densities the
Fermi energy is in excess of tens of MeV,  neutron stars are
largely unaffected by thermal effects, and the matter in their interior
exhibits the properties of cold matter at extremely high densities,
very far from those realized in  terrestrial experiments. In the era of
multi-messenger astronomical observations neutron stars offer a unique
opportunity to test a broad class of theories, from nuclear physics 
to general relativity.

From the surface to the interior of a neutron star (NS) stellar matter
undergoes a number of transitions.  From electrons and neutron rich
ions in the outer envelopes, the composition is supposed  to change
to a degenerate gas of neutrons, protons, electrons and muons in the
outer core. The composition of neutron stars in the inner core is still
to be understood. However, one possible scenario is that at densities 
larger than few times $\rho_0$ new hadronic degrees of freedom, or 
more exotic phases, are likely to appear.

The knowledge of the equation of state (EOS) of pure neutron matter
is an important bridge between the symmetry energy and
neutron star properties.  The symmetry energy $E_{\rm sym}$ is the difference of
nuclear matter and neutron matter energy and it gives the energy cost of the
isospin-asymmetry in the homogeneous nucleonic matter. In the last few
years the study of $E_{\rm sym}$ has received considerable attention (see for example
Ref.~\cite{Tsang:2012} for a recent experimental/theoretical review).

The role of the symmetry energy is
essential to understand the mechanism of stability of very-neutron
rich nuclei, but it is also related to many phenomena occurring in
neutron stars.
The stability of matter inside neutron stars is sensitive to
$E_{\rm sym}$ and its first derivative.
Neutrons tend to decay to protons through the
$\beta$-decay, and the cooling of neutron stars is related to
the proton/neutron ratio as a function of the density.  This ratio is
mainly governed by the behavior of $E_{\rm sym}$ as a function of the density.

The inner crust of neutron stars, where the density is a fraction
of nuclear density, is mostly composed of neutrons surrounding
a matter made of extremely-neutron rich nuclei that, depending
on the density, may exhibit very different phases and properties.
The extremely rich phase diagram of the neutron crustal matter is
also related to $E_{\rm sym}$.  For example it governs the
phase-transition between the crust and the core~\cite{Steiner:2015}
and $r$-mode instability~\cite{Wen:2012,Vidana:2012}.

The calculation of the EOS of neutron matter is particularly difficult 
because neutron
matter is one of the most strongly-correlated fermionic systems.
Neutron matter is often modeled by density functionals. Traditional
Skyrme models (see for example Ref.~\cite{Stone:2007} and references
therein) and relativistic mean-field models (see for example
Refs.~\cite{Fattoyev:2010,Fattoyev:2013}) are two general classes of
density functional theories. 
Another class of these calculations uses nuclear potentials,
like Argonne and Urbana/Illinois forces, that reproduces
two-body scattering and properties of light nuclei with very high
precision~\cite{Carlson:2015}. For a recent review of neutron matter
see Ref.~\cite{Gandolfi:2015}.

In this paper we summarized results of the equation of state of pure neutron 
and $\Lambda$-neutron matter based on quantum Monte Carlo (QMC) methods, that
are used to accurately calculate properties of nuclear systems in 
a non-perturbative framework.

\section{Nuclear Hamiltonians and quantum Monte Carlo methods}

In our model, nuclei and neutron matter are described by non-relativistic point-like particles
interacting via two- and three-body forces:
\begin{align}
	H_{\rm nuc} = \sum_i\frac{p_i^2}{2m_N} + \sum_{i<j}v_{ij} + \!\!\sum_{i<j<k}v_{ijk} \;.
\end{align}
The two body-potential that we use is the Argonne
AV8'~\cite{Wiringa:2002}, that is a simplified form of the Argonne
AV18~\cite{Wiringa:1995}. Although simpler to use in QMC calculations,
the AV8' provides almost the same accuracy as AV18 in fitting NN
scattering data.  The three-body force is not as well constrained
as the NN interaction, but its inclusion in realistic nuclear
Hamiltonians is important to correctly describe the binding energy
of light nuclei~\cite{Carlson:2015}. 
The Urbana IX (UIX) three-body force has been originally proposed in
combination with the Argonne AV18 and AV8'~\cite{Pudliner:1995}. Although
it slightly underbinds the energy of light nuclei, it has been
extensively used to study the equation of state of nuclear and neutron
matter~\cite{Akmal:1998,Gandolfi:2009,Gandolfi:2012}. 
In this paper we
shall present a study of the neutron matter EOS based on different models
of three-neutron forces giving specific values of the symmetry 
energy~\cite{Gandolfi:2012,Gandolfi:2014}.

For $\Lambda$-hypernuclei and $\Lambda$-neutron matter the Hamiltonian
is modified as 
\begin{align}
	H_{\rm hyp} = H_{\rm nuc} + \sum_{\lambda}\frac{p_\lambda^2}{2m_\Lambda} + \sum_{\lambda i}v_{\lambda i}
	+ \!\!\sum_{\lambda,i<j}v_{\lambda ij}\;,
\end{align}
where latin indices $i,j$ label nucleons and the greek symbol $\lambda$ is used
for $\Lambda$~particles. In the strange sector we adopt explicit 
$\Lambda$N and $\Lambda$NN phenomenological interactions analogous 
to the Argonne-Illinois nucleon-nucleon force~\cite{Usmani:1995,Usmani:1995_3B,Usmani:2008,Imran:2014}.

In the non-strange sector we employ a simplified interaction
in order to make the calculations feasible also for heavier
hypernuclei. In particular we use Argonne AV4' two-body 
interaction~\cite{Wiringa:2002} plus the central repulsive term
of the three-body Urbana IX potential~\cite{Pudliner:1997}. This
choice provides a realistic description of closed shell nuclei up to
$A=48$~\cite{Pederiva:2015}. 

The two-body $\Lambda$N force is modeled with a Urbana-type potential~\cite{Lagaris:1981}, 
consistent with the available $\Lambda p$ scattering data
\begin{align}
	v_{\lambda i}=v_{0}(r_{\lambda i})+\frac{1}{4}v_\sigma T^2_\pi(r_{\lambda i})\,{\bm\sigma}_\lambda\cdot{\bm\sigma}_i \;,
	\label{eq:V_LN}
\end{align}
while the three-body potential $v_{\lambda ij}$ is written as the sum of 2$\pi$-exchange contributions
$v^{{2\pi}}_{\lambda ij}=v^{2\pi,P}_{\lambda ij}+v^{2\pi,S}_{\lambda ij}$ and a spin-dependent dispersive term 
$v_{\lambda ij}^{D}$:
\begin{align}
	v_{\lambda ij}^{2\pi,P}&=-\frac{C_P}{6}
	\Bigl\{X_{i\lambda}\,,X_{\lambda j}\Bigr\}\,{\bm\tau}_{i}\cdot{\bm\tau}_{j}\;,\nonumber\\[0.7em]
	v_{\lambda ij}^{2\pi,S}&=
	C_S\,Z\left(r_{\lambda i}\right)Z\left(r_{\lambda j}\right)\,
	{\bm\sigma}_{i}\cdot\hat{\bm r}_{i\lambda}\,
	{\bm\sigma}_{j}\cdot\hat{\bm r}_{j\lambda}\,{\bm\tau}_{i}\cdot{\bm\tau}_{j}\;,\label{eq:V_LNN}\\[0.7em]
	v_{\lambda ij}^{D}&=W_D\,
	T_{\pi}^{2}\left(r_{\lambda i}\right)T^{2}_{\pi}\left(r_{\lambda j}\right)
	\bigg[1+\frac{1}{6}{\bm\sigma}_\lambda\cdot\left({\bm\sigma}_{i}+{\bm\sigma}_{j}\right)\bigg]\;.\nonumber
\end{align}
All the details of the hypernuclear interaction, together with
the complete list of parameters, can be found in
Refs.~\cite{Usmani:2008,Imran:2014,Lonardoni:2014}.

We solve the many-body ground-state using the auxiliary field
diffusion Monte Carlo (AFDMC) originally introduced by Schmidt and
Fantoni~\cite{Schmidt:1999}. The main idea of QMC methods is to evolve
a many-body wave function in imaginary-time:
\begin{align}
	\Psi(\tau)=\exp\big[-H\tau\big]\,\Psi_v \;,
\end{align}
where $\Psi_v$ is a variational ansatz and $H$ is the Hamiltonian
of the system. In the limit of $\tau\rightarrow\infty$, $\Psi$
approaches the ground-state of $H$. The evolution in imaginary-time
is performed by sampling configurations of the system using Monte
Carlo techniques, and expectation values are evaluated over
the sampled configurations. For more details see for example
Refs.~\cite{Pudliner:1997,Gandolfi:2009,Carlson:2015}.

The Green's Function Monte Carlo (GFMC) method is extremely accurate in the
study of properties of light nuclei. The variational wave function
includes all the possible spin/isospin states of nucleons and it
provides a good variational ansatz to start the projection in the
imaginary-time. The exponential growing of this states limits the
calculation to $^{12}$C~\cite{Pieper:2005,Lovato:2013}.
The AFDMC method does not explicitly include all the spin/isospin states in the wave
function, but they are instead sampled using the Hubbard-Stratonovich
transformation. The calculation can be then extended up to many neutrons,
making the simulation of homogeneous matter possible. The AFDMC has proven to
be very accurate when compared to GFMC calculation of energies of neutrons
confined in an external potential~\cite{Gandolfi:2011}.

\section{The equation of state at nuclear densities and the symmetry energy}

The symmetry energy is defined as the energy difference between
pure neutron matter and symmetric nuclear matter. The energy of nuclear
matter is often expressed as an expansion in even powers of the
isospin-asymmetry
\begin{align}
	E(\rho,x)=E_0(\rho)+E_{\rm sym}^{(2)}(\rho)(1-2x)^2+E_{\rm sym}^{(4)}(1-2x)^4+\dots \;,
\end{align}
where $E$ is the energy per particle, $x=\rho_p/(\rho_p+\rho_n)$ is
the proton fraction, $\rho$ is the density of the system, $E_{\rm sym}^{(2n)}$
are coefficients multiplying the isospin asymmetry terms $(1-2x)^{2n}$, and
$E_0(\rho)=E(\rho,x=0.5)$ is the energy of symmetric nuclear matter.
The symmetry energy $E_{\rm sym}$ is given by
\begin{align}
	E_{\rm sym}(\rho)=E(\rho,0)-E_0(\rho) \;.
\end{align}
The energy of symmetric nuclear matter at saturation extrapolated
from the binding energy of heavy nuclei is $E(\rho_0)=-16$~MeV. 
The symmetry energy around saturation $\rho_0$ can be expanded as
\begin{align}
	E_{\rm sym}(\rho)\Big|_{\rho_0}=E_{\rm sym}+\frac{L}{3}\frac{\rho-\rho_0}{\rho_0}+\dots \;,
	\label{eq:lvsesym}
\end{align}
where $L$ is related to the slope of $E_{\rm sym}$.
By combining the above equations, we can easily relate the symmetry
energy to the EOS of pure neutron matter at density close to $\rho_0$.

\begin{figure}[h!]
 	\centering
	\includegraphics[width=0.7\textwidth]{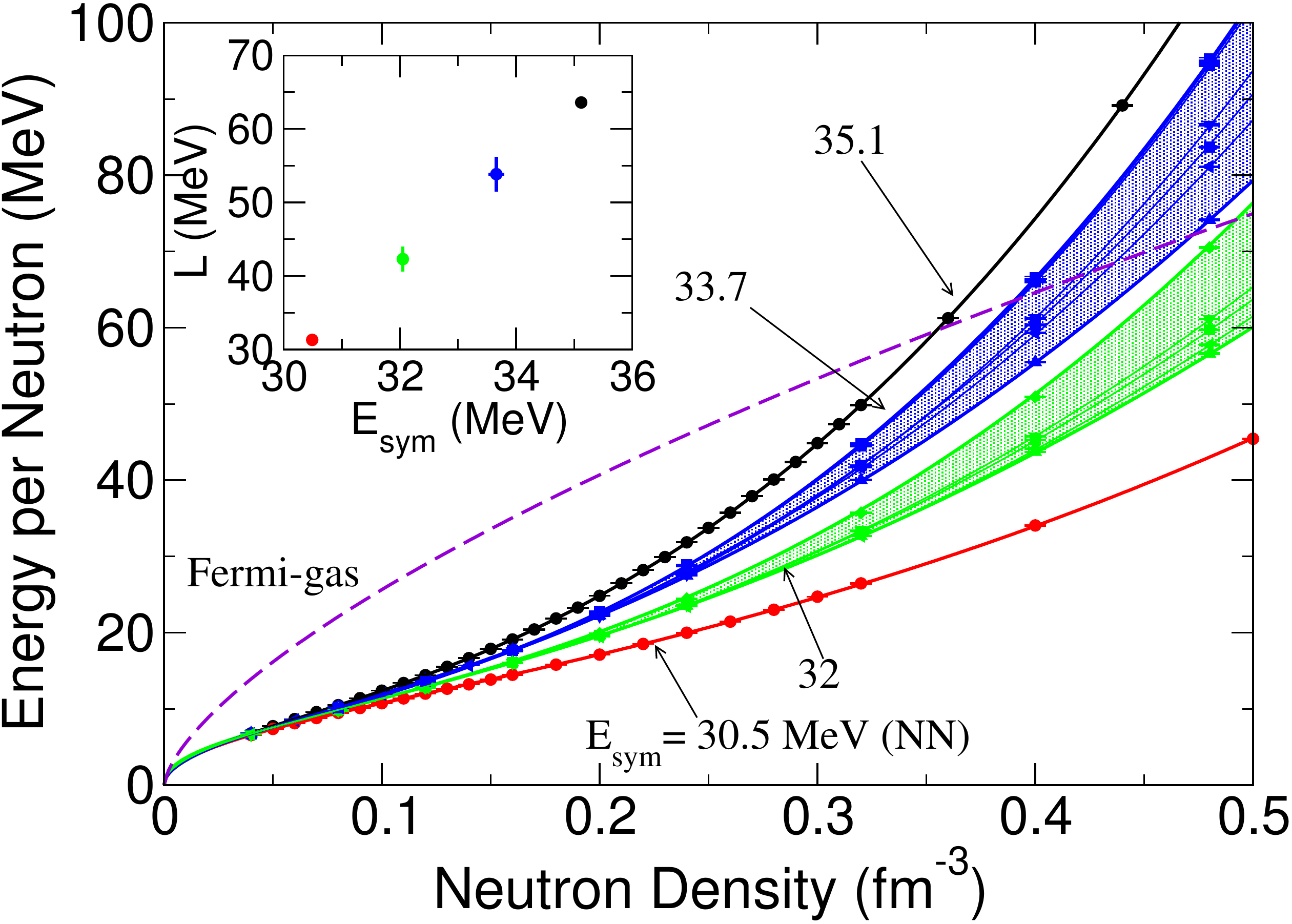}
	\caption[]{The QMC equation of state of neutron matter for various Hamiltonians.
	The red (lower) curve is obtained by including the NN (Argonne AV8') alone in the
	calculation, and the black one is obtained by adding the Urbana IX three-body force.
	The green and blue bands correspond to EOSs giving the same $E_{\rm sym}$ (32 and 33.7~MeV
	respectively), and are obtained by using several models of three-neutron force.
	In the inset we show the value of $L$ as a function of $E_{\rm sym}$ obtained by fitting
	the EOS.
	The figure is taken from Ref.~\cite{Gandolfi:2012}.}
	\label{fig:eos}
\end{figure}

We present several EOSs obtained using different models of three-neutron
force in Fig.~\ref{fig:eos}. The two solid lines correspond to the EOSs
calculated using the NN potential alone and including the UIX three-body
force. The effect of using different models of three-neutron force
is clear in the two bands, where the high density behavior is shown up
to about $3\rho_0$.  At such high density, the various models giving the
same symmetry energy at saturation produce an uncertainty in the EOS of
about 20~MeV.
The EOS obtained using QMC can be conveniently fit using the following
functional~\cite{Gandolfi:2009}:
\begin{align}
	E(\rho)=a\,\left(\frac{\rho}{\rho_0}\right)^\alpha
	+b\,\left(\frac{\rho}{\rho_0}\right)^\beta \;,
	\label{eq:enefunc}
\end{align}
where $E$ is the energy per neutron,
and $a$, $b$, $\alpha$ and $\beta$ are free parameters.
The parametrizations of the EOS obtained from different nuclear Hamiltonians
is given in Refs.~\cite{Gandolfi:2012,Gandolfi:2014}.

At $\rho_0$ symmetric nuclear matter saturates, and we can extract the value of $E_{\rm sym}$ and $L$
directly from the pure neutron matter EOS. The result of fitting
Eq.~(\ref{eq:lvsesym}) to the pure neutron matter EOS is shown in the
inset of Fig.~\ref{fig:eos}.  The error bars are obtained by taking the
maximum and minimum value of $L$ for a given $E_{\rm sym}$, and the curves
obtained with NN and NN+UIX are thus without error bars. From the plot it
is clear that within the models we consider, the correlation between $L$
and $E_{\rm sym}$ is linear and quite strong. This conclusion is even more evident
in Ref.~\cite{Gandolfi:2014} where more different forms of three-body forces have
been considered.

\section{Neutron star structure}
When the EOS of the neutron matter has been specified, the structure
of an idealized spherically-symmetric neutron star model can be calculated
by integrating the Tolman-Oppenheimer-Volkoff (TOV) equations:
\begin{align}
	\frac{dP}{dr}=-\frac{G[m(r)+4\pi r^3P/c^2][\epsilon+P/c^2]}{r[r-2Gm(r)/c^2]} \;,
	\label{eq:tov1}
\end{align}
\begin{align}
	\frac{dm(r)}{dr}=4\pi\epsilon r^2 \;,
	\label{eq:tov2}
\end{align}
where $P=\rho^2(\partial E/\partial\rho)$ and $\epsilon=\rho(E+m_N)$
are the pressure and the energy density, $m_N$ is the neutron mass,
$m(r)$ is the gravitational mass enclosed within a radius $r$, and $G$
is the gravitational constant. The solution of the TOV equations for a
given central density gives the profiles of $\rho$, $\epsilon$ and $P$ as
functions of radius $r$, and also the total radius $R$ and mass $M=m(R)$.
The total radius $R$ is given by the condition $P(R)=0$.

\begin{figure}[h!]
 	\centering
	\includegraphics[width=0.7\textwidth]{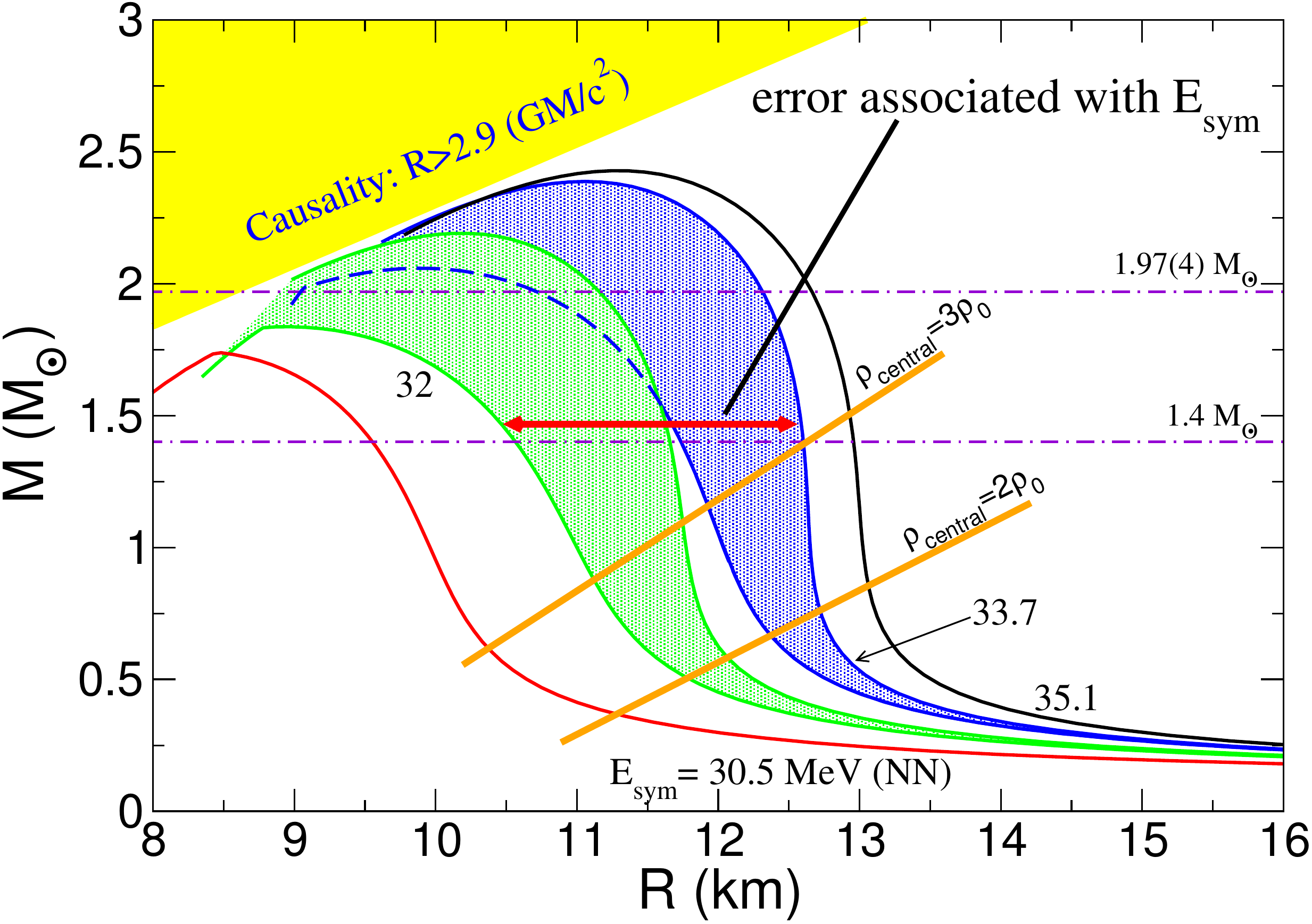}
	\caption[]{The mass-radius relation of neutron stars obtained from
	the EOS calculated using QMC. The various colors represent the $M-R$
	result obtained from the corresponding EOSs described in
	Fig.~\ref{fig:eos}. The two horizontal lines show the value of $M=1.4 M_{\odot}$
	and $1.97(4) M_{\odot}$~\cite{Demorest:2010}.
	The figure is taken from Ref.~\cite{Gandolfi:2012}.}
	\label{fig:nstar}
\end{figure}

The mass of a neutron star as a function of its radius is shown in
Fig.~\ref{fig:nstar}. The two bands correspond to the result obtained
using the two sets of EOS giving the same value of $E_{\rm sym}$ indicated
in the figure. As in the case of the EOS, it is clear that the main
source of uncertainty in the radius of a neutron star with $M=1.4M_{\odot}$
is due to the uncertainty on $E_{\rm sym}$ rather than the model of
the three-neutron force.
It has to be noted that we have used the EOS of pure neutron matter
without imposing the $\beta$-equilibrium, so in our model we do not have
protons. However, the addition of a small proton fraction would only slightly change
the radius $R$~\cite{Gandolfi:2010,Akmal:1998}, resulting in a difference smaller
than other uncertainties in the EOS that we have discussed.

The EOS of neutron matter and its properties can also be
extracted from astrophysical observations~\cite{Steiner:2010}. By
combining the Bayesian analysis with the model of neutron matter of
Eq.~(\ref{eq:enefunc}) it is possible to compare the QMC prediction with
observations~\cite{Steiner:2012} and to extract $E_{\rm sym}$ and $L$:
\begin{align}
	E_{\rm sym}=a+b+16 \;,\quad\quad L=3\,(a\alpha+b\beta) \;.
\end{align}
From neutron stars we obtain the constraints
$31.2~{\rm MeV}<E_{\rm sym}<34.3~{\rm MeV}$ and 
$36.6~{\rm MeV}<L<55.1~{\rm MeV}$~\cite{Steiner:2012} at the
$2$-$\sigma$ confidence level, in agreement with QMC predictions.

\begin{figure}[h!]
 	\centering
	\includegraphics[width=0.7\textwidth]{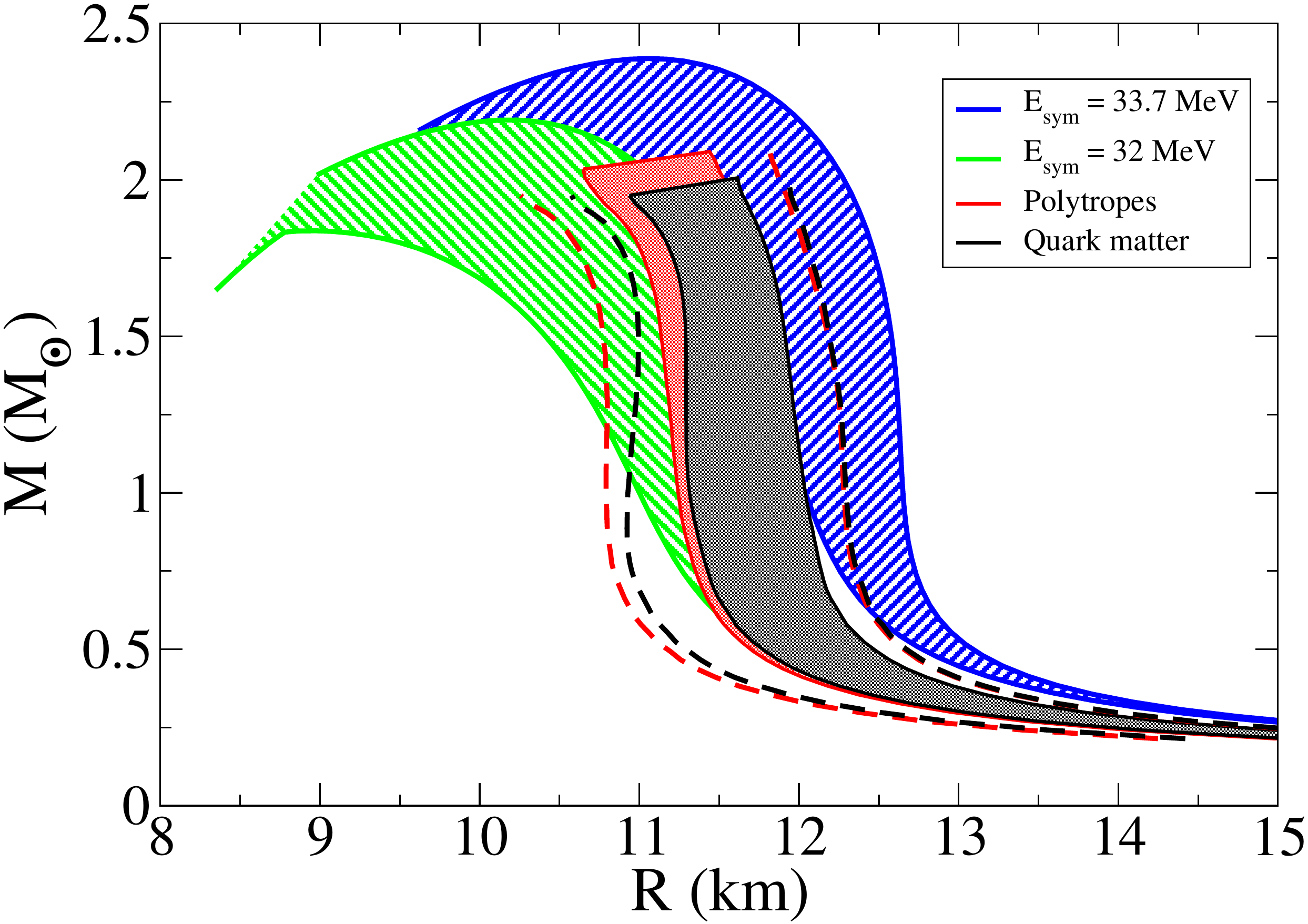}
	\caption[]{The comparison of the $M-R$ relation of neutron stars obtained
	from QMC calculations and observations.
	The blue and green bands are the same as Fig.~\ref{fig:nstar} and
	correspond to EOSs giving the value of $E_{\rm sym}$ indicated in the legend.
	The black and red bands are obtained from neutron star observations of
	Ref.~\cite{Steiner:2012} at the $1$-$\sigma$ confidence level (dashed lines
	at $2$-$\sigma$), and they correspond to different models
	of the high-density EOS.}
	\label{fig:mvsr}
\end{figure}

The agreement between theoretical calculations with the neutron star structure obtained 
from observations is well represented in Fig.~\ref{fig:mvsr}. The two
green and blue bands correspond to the $M-R$ relation obtained from the
EOS of Fig.~\ref{fig:eos}, and the black and red bands represent the
astrophysical observation of Ref.~\cite{Steiner:2012} using different
models for the high-density EOS.

\section{Neutron matter at high density}

In Ref.~\cite{Ambartsumyan:1960} Ambartsumyan and
Saakyan reported the first theoretical indication of the appearance
of hyperons in the core of a NS.  In the degenerate dense matter of
the inner core, when the nucleon chemical potential is large enough,
the conversion of nucleons into hyperons might become energetically
favorable. On the other hand, Pauli blocking would prevent hyperons
from decaying by limiting the phase  space available to nucleons.
This would lead to a reduction of the Fermi pressure exerted by
the baryons and to a softening of the EOS, and, as a
consequence, the predicted maximum mass of neutron stars would be reduced.
However, the recent measurements of the large neutron star mass values 
of $1.97(4)M_\odot$~\cite{Demorest:2010} and $2.01(4)M_\odot$~\cite{Antoniadis:2013}
require a stiff equation of state. Other NS observations of
masses and radii seem to disfavor a very soft EOS of neutron star 
matter~\cite{Steiner:2010,Steiner:2012,Bedaque:2015}.  This
seems to contradict the appearance of strange baryons in high-density
matter, given what is known at present about  the hyperon-nucleon
interaction. This apparent inconsistency between NS mass observations
and theoretical calculations  is a long standing problem known as
\emph{hyperon puzzle}. 

Currently there is no general agreement (even qualitative) among
the predicted results for the EOS and the  maximum mass of a NS
including hyperons. This has to be ascribed to the combination 
of an incomplete knowledge of the forces governing the system (in
the hypernuclear case both two- and three-body ones),  and to the
concurrent use of approximated theoretical many-body techniques.
Some classes of methods extended to the hyperonic sector predict
the appearance of hyperons at around $2-3\rho_0$,  and a strong
softening of EOS, implying a sizable reduction of the maximum
mass~\cite{Dapo:2010,Schulze:2011,Vidana:2011,Massot:2012}. 
On the other hand, other approaches suggest much weaker
effects arising from the presence of strange baryons in the core of the 
star~\cite{Bednarek:2012,Weissenborn:2012,Jiang:2012,Miyatsu:2013,Gupta:2013,Yamamoto:2014}.

The large body of available nucleon-nucleon scattering
data allows to derive satisfactory
models of  two-body nuclear forces, either purely
phenomenological or built on the basis of an effective field 
theory~\cite{Wiringa:1995,Machleidt:1996,Gezerlis:2013}.
In the hyperon-nucleon sector, several models of two-body
force are available~\cite{Rijken:2006,Rijken:2010,Haidenbauer:2013}, 
but they relies on a poor experimental knowledge.
Few hyperon-nucleon scattering data are available, and  no scattering
data exist in the hyperon-hyperon sector. The main reasons of this
lack of information lie in the instability  of hyperons in the vacuum,
and the impossibility of collecting hyperon-neutron and hyperon-hyperon
scattering data. This implies that realistic hypernuclear interaction
models must also rely on information extracted from the binding energies
of hypernuclei.

\begin{figure}[b]
 	\centering
	\includegraphics[width=0.7\textwidth]{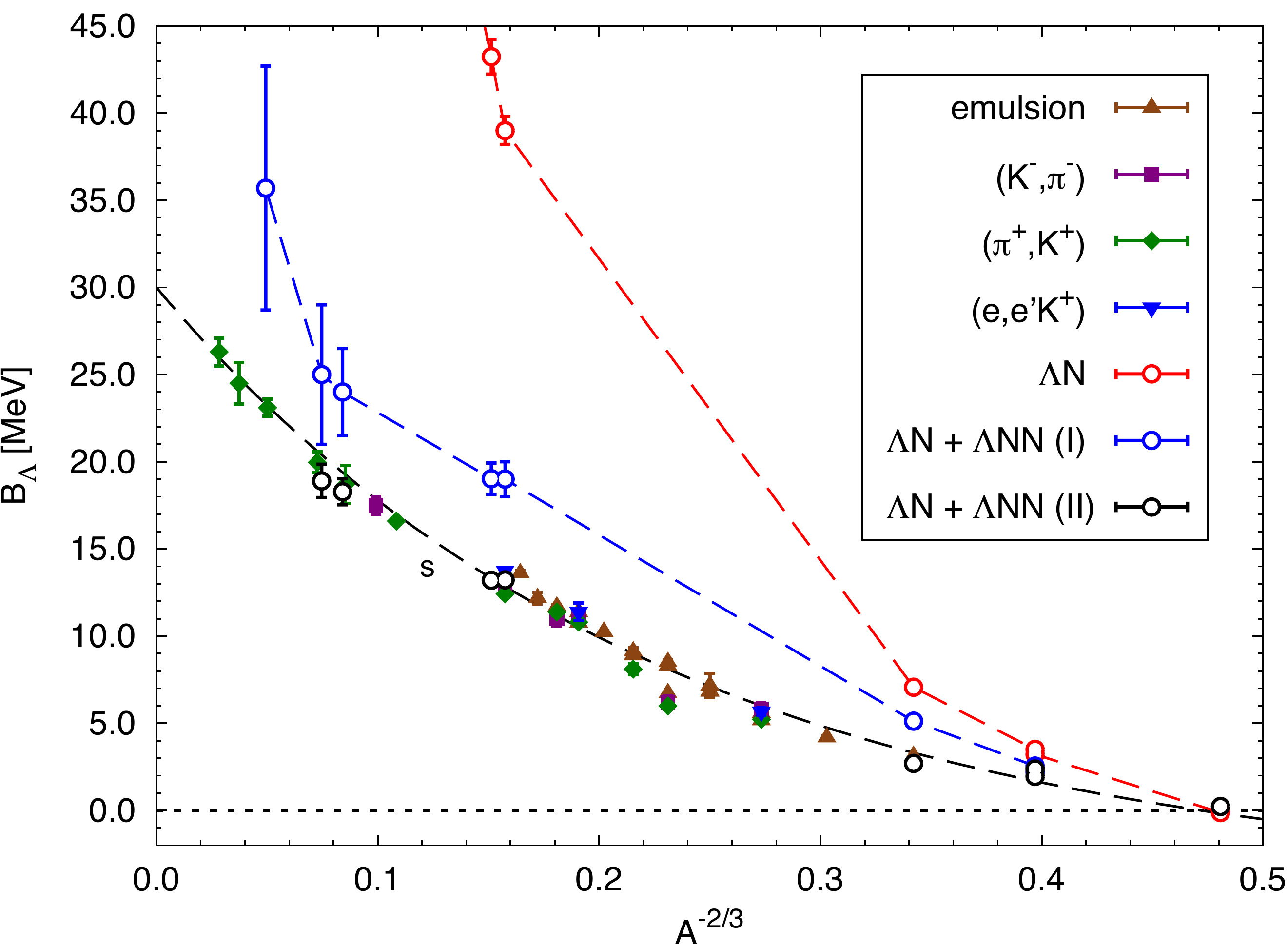}
	\caption[]{Solid symbols are the available $B_\Lambda$ experimental values
	in $s$ wave for different hypernuclear production mechanisms 
	(see Ref.~\cite{Pederiva:2015} for the complete list of experimental references). 
	Empty symbols refer to quantum Monte Carlo results. Red dots (upper curve) is the case of two-body $\Lambda$N interaction alone. 
	Blue (middle curve) and black (lower curve) dots are the results obtained 
	including two different parametrizations of the three-body hyperon-nucleon force~\cite{Lonardoni:2014,Pederiva:2015}.}
	\label{fig:bl1} 
\end{figure}

\begin{figure}[h!]
	\centering
	\includegraphics[width=0.7\textwidth]{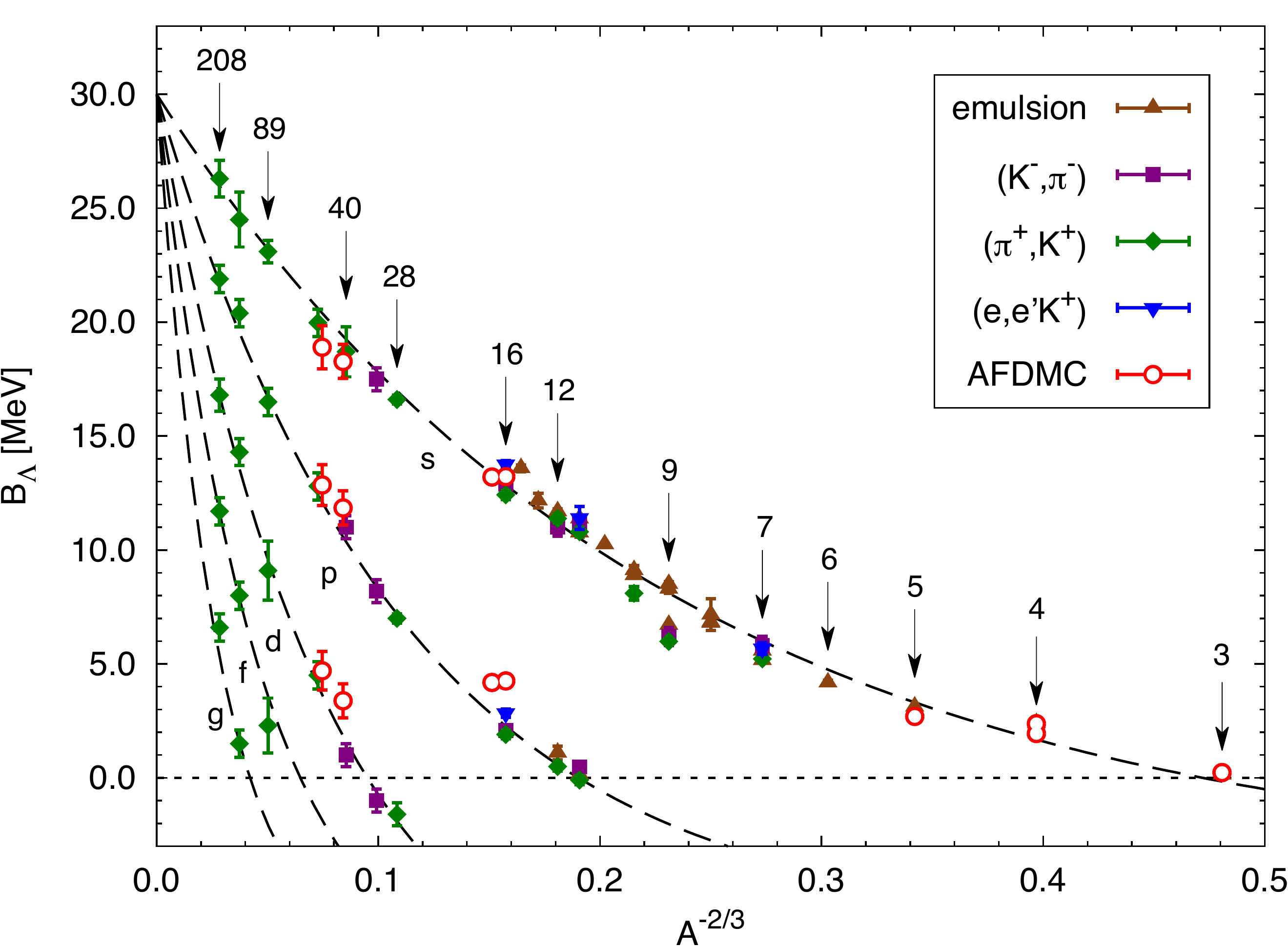}
	\caption[]{Solid symbols refer to the experimental results for the $\Lambda$ particle in $s$, $p$, $d$, $f$ and $g$ waves.
	Red empty dots are the quantum Monte Carlo results obtained including the most recent two- plus three-body hyperon-nucleon
	phenomenological interaction model~\cite{Pederiva:2015}.}
	\label{fig:bl2} 
\end{figure}

In Refs.~\cite{Lonardoni:2013,Lonardoni:2014,Pederiva:2015} it has
been shown that the repulsive nature of the three-body hyperon-nucleon
interaction is the key to satisfactorily reproduce the ground state
properties of light- to medium-heavy hypernuclei within a unique
theoretical framework. By means of a re-fit of the $\Lambda$NN force
of Eq.~(\ref{eq:V_LNN}) to the available $\Lambda$~separation energies
of closed-shell $\Lambda$-hypernuclei, AFDMC calculations result
in a good agreement with experimental data over a wide mass range.
In Fig.~\ref{fig:bl1} we show the binding energy of several hypernuclei
calculated with AFDMC and compared with a selection of experimental data. 
The AFDMC results have been obtained by including the two-body $\Lambda$N force 
alone (dashed red upper curve), and together with two models of $\Lambda$NN interaction,
$\Lambda$NN(I) initially proposed by Usmani~\cite{Usmani:1995}, and $\Lambda$NN(II)
of Ref.~\cite{Lonardoni:2014}. Hypernuclear three-body forces are crucial to 
qualitatively and quantitatively reproduce the binding energy of hypernuclei. 
More results, including several excited
states of the hyperon, are reported in Fig.~\ref{fig:bl2}~\cite{Pederiva:2015}.

However, parametrizations of the potential predicting relatively small
differences in the $\Lambda$ separation energies of hypernuclei 
give very different results for the properties of the infinite medium
\cite{Lonardoni:2015}.

We define the total baryon density $\rho$ and the $\Lambda$~fraction $x$ as:
\begin{align}
	\rho=\rho_n+\rho_\Lambda \;, \quad\quad
	x=\frac{\rho_\Lambda}{\rho} \;,
\end{align}
where $\rho_n$ and $\rho_\Lambda$ are the neutron and $\Lambda$ densities.
The total energy of the $\Lambda$-neutron matter is given by
\begin{align}
	E_{\rm{\Lambda nm}}(\rho,x)=\left[E_{\rm{pnm}}((1-x)\rho)+m_N\right](1-x)
	+\left[E_{\rm{p\Lambda m}}(x\rho)+m_\Lambda\right]x+f(\rho,x)\;,
\end{align}
where $E_{\rm{pnm}}$ is the one defined in Eq.~(\ref{eq:enefunc}), $E_{\rm p\Lambda m}$ 
is the non-interacting energy of pure $\Lambda$ matter, and the $\Lambda$-neutron
part is parametrized as
\begin{align}
	f(\rho,x)=c_1\frac{x\,(1-x)\,\rho}{\rho_0}+c_2\frac{x\,(1-x)^2\,\rho^2}{\rho_0^2} \;.
\end{align}
The two coefficients $c_1$ and $c_2$ have been obtained by fitting AFDMC results
of $\Lambda$-neutron matter at various densities $\rho$ and concentrations $x$.
At each density, the fraction $x$ as a function of $\rho$ is obtained by 
imposing chemical equilibrium, and finally, for a given Hamiltonian, an EOS 
containing neutrons and $\Lambda$s in chemical equilibrium is obtained.

\begin{figure}[h!]
	\centering
	\includegraphics[width=0.7\textwidth]{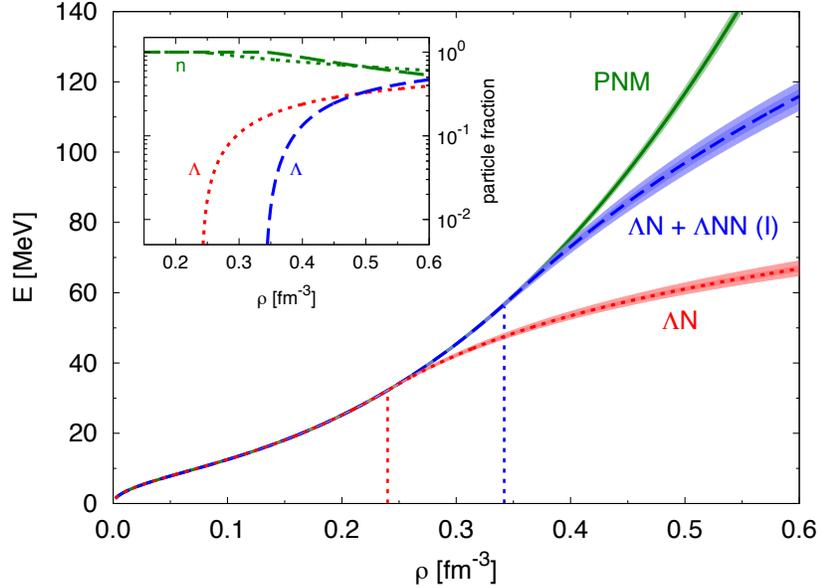}
	\caption[]{Equations of state. Green solid curves refer to pure neutron matter 
	calculated with realistic two- plus three-nucleon potentials. 
	The red dotted curve represents hyper matter with hyperons interacting via the two-body $\Lambda$N force alone. 
	The blue dashed and black dotted-dashed curves are obtained including two different parametrizations of the 
	three-body hyperon-nucleon potential. Shaded regions represent the uncertainties on the results. 
	In the inset, neutron and lambda fractions corresponding to the two hyper-neutron matter EOSs.
	The figure is taken from Ref.~\cite{Lonardoni:2015}.}
	\label{fig:eoshyp} 
\end{figure}

The resulting EOSs span the whole regime 
extending from the appearance of a substantial fraction of hyperons
at $\sim2\rho_0$ to the absence of $\Lambda$ particles in the entire 
density range of the star. In Fig.~\ref{fig:eoshyp} we show the AFDMC results
obtained for pure neutron matter (same reported in Fig.~\ref{fig:eos}), and 
with the addition of $\Lambda$ particles interacting with the hypernuclear forces 
employed for hypernuclei (results reported in Fig.~\ref{fig:bl1}).
In the inset the neutron and $\Lambda$ fractions are shown. Note that
the $\Lambda$NN interaction that better reproduces the binding energy of hypernuclei,
$\Lambda$NN(II) in Fig.~\ref{fig:bl1} and used in Fig.~\ref{fig:bl2}, 
gives a zero $\Lambda$ fraction at least up to $\rho=0.56~\text{fm}^{-3}$,
and then the EOS corresponds to the pure neutron matter one up to such a density.

\begin{figure}[h!]
	\centering
	\includegraphics[width=0.7\textwidth]{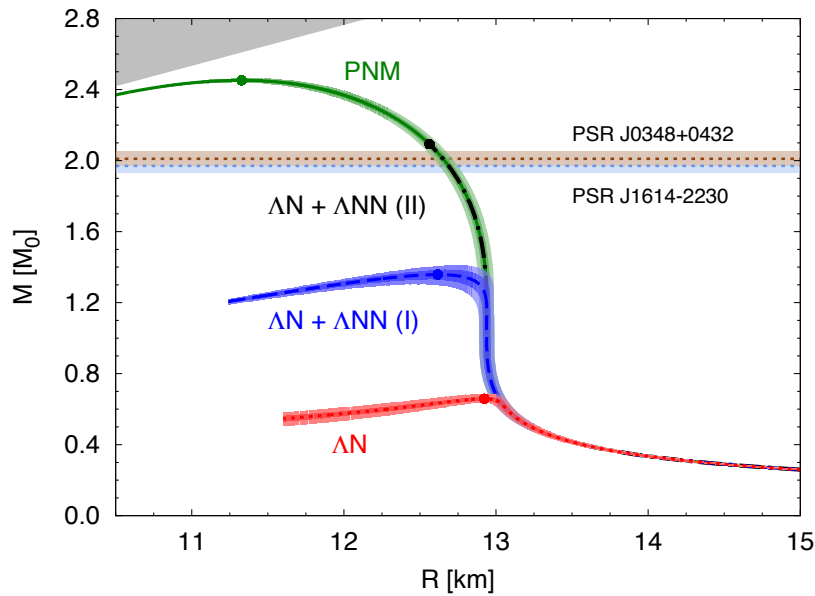}
	\caption[]{Mass-radius relations. The color scheme is the same as Fig.~\ref{fig:eoshyp}. 
	Full dots represent the predicted maximum masses. 
	Horizontal bands at $\sim 2M_\odot$ are the observed masses of the heavy pulsars PSR~J1614-2230~\cite{Demorest:2010} and
	PSR~J0348+0432~\cite{Antoniadis:2013}. The grey shaded region is the excluded part of the plot due to causality.
	The figure is taken from Ref.~\cite{Lonardoni:2015}.}
	\label{fig:mofr} 
\end{figure}

As suggested by the qualitative change of the EOS, the addition of $\Lambda$ hyperons
to neutron matter yields to a sizable effect on the predicted neutron star structure
(see Fig.~\ref{fig:mofr}).
The two Hamiltonians that overbind hypernuclei, i.e. $\Lambda$N and 
$\Lambda{\rm N}+\Lambda{\rm NN}$(I), produce too low neutron star masses.
In particular, in the latter case hyperons appear at around twice 
saturation density and the predicted maximum mass is less than $1.4M_\odot$.
The EOS for the Hamiltonian $\Lambda{\rm N}+\Lambda{\rm NN}$(II) is instead stiff 
enough to support the observations with a lower limit for the predicted maximum mass 
of $2.09(1)M_\odot$.

These results suggest that within the $\Lambda$N model that we have considered, the presence
of hyperons in the core of the neutron stars cannot be satisfactorily
established, and thus there is no clear incompatibility with astrophysical
observations when lambdas are included. Therefore, the derivation of
realistic hypernuclear potential models is of primary importance to
properly assess the role of hyperons to the neutron star structure.
This demands a precise and systematic experimental investigation of 
properties of hypernuclei over
a wide range of masses. In this direction a recent study of the isospin
dependence of the  present three-body hyperon-nucleon force has been
carried out, underlying the difficulties in extracting the information
on the Hamiltonian  from currently available experimental information
on hypernuclei~\cite{Pederiva:2015}.

\section{Conclusions}
Quantum Monte Carlo calculations have been extensively used to derive
properties of neutron matter at different density regimes to
study neutron star structure.
The neutron matter EOS around saturation mostly determines the radii of neutron
stars~\cite{Lattimer:2001}, and it is directly related to the symmetry energy.
We have calculated the EOS, and quantified the connection between $E_{\rm sym}$ and
radii of neutron stars. 
Although the EOS of neutron matter is qualitatively understood around nuclear
density, the appearance of hyperons in the inner core of neutron stars
strongly affects the prediction of neutron star properties like
the maximum mass.
However, by employing hypernuclear interactions that successfully reproduce the experimental 
separation energies of hypernuclei, the presence of hyperons in the
core of the neutron stars cannot be satisfactorily established.
The hyperon-neutron force will need both additional theoretical 
investigation and a substantial additional amount of experimental data, 
in particular for highly asymmetric hypernuclei and 
excited states of the hyperon.

\section{Acknowledgement}
We would like to thank J. Carlson, F. Catalano, A. Lovato, F. Pederiva, S.~C. Pieper, S. Reddy, A.~W. Steiner,
W. Weise, and R.~B. Wiringa for
stimulating discussions and for sharing their results.
This work was supported by the U.S. Department of Energy, Office
of Science, Office of Nuclear Physics, under the NUCLEI SciDAC grant
(S.G. and D.L.) and by DOE under Contract No. DE-AC52-06NA25396 and Los
Alamos LDRD grant (S.G.). This research used resources of the National
Energy Research Scientific Computing Center (NERSC), which is supported
by the Office of Science of the U.S. Department of Energy under Contract
No. DE-AC02-05CH11231, and computing time provided by Institutional
Computing (IC) at LANL.

\end{document}